# Unusual Intralayer Ferromagnetism Between S = 5/2 ions in MnBi$_2$Te$_4$: Role of Empty Bi $p$ States

Jing Li[1,2], J. Y. Ni[1,2], X. Y. Li[1,2], H.-J. Koo[3], M.-H. Whangbo[4,5], J.S. Feng[1,6,*] and H. J. Xiang[1,2,†]

[1]*Key Laboratory of Computational Physical Sciences (Ministry of Education), State Key Laboratory of Surface Physics, and Department of Physics, Fudan University, Shanghai 200433, P. R. China*

[2]*Collaborative Innovation Center of Advanced Microstructures, Nanjing 210093, P. R. China*

[3]*Department of Chemistry and Research Institute for Basic Sciences, Kyung Hee University, Seoul 02447, Korea*

[4]*Department of Chemistry, North Carolina State University, Raleigh, North Carolina 27695-8204, United States*

[5]*Group SDeng, State Key Laboratory of Structural Chemistry, Fujian Institute of Research on the Structure of Matter (FJIRSM), Chinese Academy of Sciences (CAS), Fuzhou 350002, China*

[6]*School of Physics and Materials Engineering, Hefei Normal University, Hefei 230601, P. R. China*

*email: fjs@hfnu.edu.cn

†email: hxiang@fudan.edu.cn

**Abstract**

The layered magnetic topological insulator MnBi$_2$Te$_4$ is a promising platform to realize the quantum anomalous Hall effect because its layers possess intrinsic ferromagnetism. However, it is not well understood why the high-spin $d^5$ magnetic ions Mn$^{2+}$ forming the Mn-Te-Mn spin exchange paths prefer ferromagnetic (FM) coupling, contrary to the prediction of the Goodenough-Kanamori rule that a TM-L-TM spin exchange, where TM and L are a transition-metal magnetic cation and a main group ligand, respectively, is antiferromagnetic (AFM) even when the bond angle of the exchange path is 90°. Using density functional theory (DFT) calculations, we show that the presence of Bi$^{3+}$ ions is essential for the FM coupling in MnBi$_2$Te$_4$. Then, using a tight-binding model Hamiltonian, we find that high-spin $d^5$ ions (S = 5/2) in TM-L-TM spin exchange paths prefer FM coupling if the empty $p$-orbitals of a nonmagnetic cation M (e.g., Bi$^{3+}$ ion) hybridize strongly with those of the bridging ligand L, but AFM coupling otherwise.

In the context of tremendous progress in topological materials [1,2], many attempts have been made to introduce magnetism in topological insulators (TI) for emerging new physics and potential applications [3-8], which include the quantum anomalous Hall (QAH) effect and topological magnetoelectric effect. It is difficult to combine magnetism and topological properties in a natural insulator simultaneously. So far, there are two dominating methods of introducing magnetism, one is doping magnetic metal elements into TI [4,9,10], and the other is constructing ferromagnet/TI heterostructure [11-15]. The former led to the experimental observation of the QAH effect at tens of millikelvin [10], and the latter has complicated technical requirements. Obviously, both are at the stage far from practical applications. Recently, the layered phase $MnBi_2Te_4$ was theoretically predicted to be an antiferromagnetic (AFM) TI [16,17] and was soon confirmed by experiments [18-20]. Each $MnBi_2Te_4$ layer has a seven-sheet structure with the stacking pattern of Te-Bi-Te-Mn-Te-Bi-Te [see FIG. 1(a)] where each sheet has a trigonal arrangement of atoms; the sheet of Mn and the two inner sheets of Te form a $MnTe_2$ layer. The latter is sandwiched between two sheets of Bi, and the resulting $MnTe_2Bi_2$ layer between two outer sheets of Te such that each Mn forms an $MnTe_6$ octahedron, and each Bi an $BiTe_6$ octahedron. The formal oxidation states of Mn, Bi and Te are +2, +3 and -2, respectively, and each $Mn^{2+}$ cation is in the high-spin state ($d^5$, S = 5/2). It is important to note that the inner Te atoms are the first-coordinate ligands of both $Mn^{2+}$ and $Bi^{3+}$ cations, namely, the inner $Te^{2-}$ anions interact with both $Mn^{2+}$ and $Bi^{3+}$ cations through the Mn-Te-Bi bridges. The $Mn^{2+}$ ions are ferromagnetically coupled in each $MnBi_2Te_4$ layer, so the topological property and magnetism of $MnBi_2Te_4$ depend on the number of $MnBi_2Te_4$ layers in a sample; $MnBi_2Te_4$ exhibits a topological axion state and antiferromagnetism in films containing an even number of layers, but the QAH effect and ferromagnetism in films containing an odd number of layers [21-24]. Due to this layer-number-dependent magnetism, $MnBi_2Te_4$ is a multi-functional magnetic material with potential applications in two dimensional (2D) magnetic materials [25,26]. Hence, it is important to study its magnetism from both a theoretical point of view and an application perspective.

Within each MnTe$_2$ layer of MnBi$_2$Te$_4$, the spin exchanges between adjacent Mn$^{2+}$ ions are Mn-Te-Mn superexchanges. In predicting whether this type of spin exchange is FM or AFM, one often employs the Goodenough-Kanamori rule [27-31]. When the TM is a high-spin $d^5$ ion, the TM-L-TM spin exchange is predicted to be strongly AFM when the TM-L-TM bond angle is around 180°, but weakly AFM when the bond angle is close to 90° [25]. Experimentally, the Mn-Te-Mn angle of MnBi$_2$Te$_4$ is 94.44° [32], so the intralayer ferromagnetism in MnBi$_2$Te$_4$ contradicts the famous Goodenough-Kanamori rule.

In this Letter, we use first-principles density functional theory (DFT) calculations to verify the ferromagnetic (FM) coupling within a MnBi$_2$Te$_4$ layer and probe the microscopic origin of the intralayer FM coupling. We find that, for the FM coupling between the Mn$^{2+}$ cations, the interactions of the Te$^{2-}$ anion in each Mn-Te-Mn spin exchange path with its adjacent Bi$^{3+}$ cation, which involves electron density transfer from the Te$^{2-}$ anion to the Bi$^{3+}$ cation, are crucial. Furthermore, using a model Hamiltonian under the tight-binding approximation, we show why this is the case.

MnBi$_2$Te$_4$ is a van der Waals (vdW) material with space group $R\bar{3}m$ for the bulk [32] and $P\bar{3}m1$ for the monolayer, and the weak vdW interactions between adjacent layers can hardly affect the intralayer magnetic interactions. Unless stated otherwise, we focus on a MnBi$_2$Te$_4$ monolayer throughout this paper. The FM coupling between magnetic ions are also found in the layered phases CrGeTe$_3$ [25] and CrI$_3$ [26] as well. However, the mechanism of FM coupling in MnBi$_2$Te$_4$ containing Mn$^{2+}$ ($d^5$, S = 5/2) ions differs completely from that in CrI$_3$ and CrGeTe$_3$, which are made up of CrI$_6$ octahedra containing Cr$^{3+}$ ($d^3$, S = 3/2) ions; in CrI$_3$ and CrGeTe$_3$, the $e_g$ state of each CrI$_6$ octahedron is empty whereas, in MnBi$_2$Te$_4$, both the $t_{2g}$ and $e_g$ states of each MnTe$_6$ octahedron are occupied. To verify the FM ground state of MnBi$_2$Te$_4$, we perform DFT+U and DFT+U+SOC (where SOC represents "spin-orbit coupling") calculations to obtain the total energies of three magnetic configurations, i.e., FM, AFM-I and AFM-II shown on FIG. 1(b) (for details of our computations, see Section I of the supporting material, SM). The energies per formula unit of the AFM-I and AFM-

II states relative to that of the FM state are summarized in FIG.1(b). In the absence and presence of SOC in the calculations, the relative energy of AFM-I is 2.42 and 4.52 meV, respectively, and that of AFM-II 3.06 and 4.48 meV, respectively. That is, regardless of whether the SOC is included or not, the ground state is always FM. Hence, it is reasonable to ignore SOC in our discussion (We will discuss briefly the effect of SOC later). Using the four-states method [33,34], we evaluate the first, second and third nearest-neighbor (NN) spin exchanges $J_1$, $J_2$ and $J_3$, respectively, to find that $J_1$ is FM (-0.94 meV), $J_2$ is AFM (0.14 meV) and $J_3$ is FM (-0.038 meV). $J_3$ is one and two orders of magnitude smaller than $J_1$ and $J_2$, respectively, and can be ignored. Because the first and second NN spin exchanges are opposite in sign, one might consider the possibility of a non-collinear spin arrangement such as a spiral order for MnBi$_2$Te$_4$. We examined this possibility by carrying out the Freiser analysis [35] (for details, see Section II and FIG. S1(a) of the SM) and Monte Carlo simulation [36] (for details, see Section III and FIG. S1(b) of the SM), to find that the ground state of MnBi$_2$Te$_4$ is indeed FM.

The crystal structure and the band dispersion relations of MnBi$_2$Te$_4$ are presented in FIG. S2, and the partial density of states (PDOS) plots in FIG. S3, in Section IV of the SM. The latter manifest that: (1) The occupied $d$-states of Mn, occurring approximately 4 eV below the Fermi level (0 eV), form a very narrow band. (2) In the occupied energy region between approximately -4 eV and the Fermi level, the $p$-states of both Te and Bi occur and overlap significantly, suggesting that there exists a strong hybridization between the Te $5p$ and Bi $6p$ states. Thus one might speculate if the hybridization of the Bi $6p$-orbitals with the $5p$-orbitals of the inner Te forming the first-coordinate of the Mn$^{2+}$ cations is responsible for the FM coupling between the Mn$^{2+}$ cations. One way to test this speculation is to do calculations for a hypothetical layer system of Mn$^{2+}$ ions that may simulate the MnTe$_2$ layer of MnBi$_2$Te$_4$ without Bi atoms. Such a system is the "MnI$_2$" layer that has the structure of the MnTe$_2$ layer of MnBi$_2$Te$_4$ (see FIG. S4 of the SM). Our calculations for this model system reveals that the AFM-I state is more stable than the FM state (by 0.41 meV per formula unit), suggesting that the Bi atoms are important for the FM coupling in MnBi$_2$Te$_4$. Nevertheless, "MnI$_2$" is different in structure from the MnTe$_2$ layer of MnBi$_2$Te$_4$. Thus, we do calculations for MnBi$_2$Te$_4$ by raising only the energy of the Bi 6$p$ orbital using the orbital selective external potential method so as to weaken the hybridization between the Bi-6$p$ and Te-5$p$ orbitals [37,38]. FIG. 2 shows that, with raising the energy level of the Bi-6$p$ orbital, the energy difference, $\Delta E = E_{AFM} - E_{FM}$, decreases almost linearly so that the AFM state becomes more stable than the FM state when the energy of the Bi-6$p$ orbital raised more than a critical value (approximately 7 eV). This result once again shows that Bi ion is necessary for the FM coupling in MnBi$_2$Te$_4$. A primary consequence of the hybridization between the Bi-6$p$ orbitals and the 5$p$ orbitals of the inner Te is that some electron density of the inner Te$^{2-}$ anions is transferred to the Bi$^{3+}$ cations. If this electron transfer is essential for the FM coupling between the Mn$^{2+}$ ions, the amount of this electron transfer should be greater in the FM than in the AFM state. As shown in Section V of the SM, this is indeed the case and the energy-lowering associated with the electron-transfer from the inner Te$^{2-}$ ions toward the Bi$^{3+}$ cations is greater in the FM state than in the AFM state.

To thoroughly understand the mechanism of the FM coupling in MnBi$_2$Te$_4$, we carry out a model Hamiltonian analysis. First, we consider a three-site cluster (TM-L-TM) model composed of two transition-metal (TM) atoms bridged by a ligand (L) atom, shown as the inset in the left panel of FIG. 3. In the case of MnBi$_2$Te$_4$, TM and L represent the Mn and Te atom, respectively. For simplicity, we consider a single orbital for each atom. There are four electrons, i.e., one for each TM atom and two for the ligand L atom. Here, we adopt tight binding (TB) method based on the mean field approximation [39,40] to obtain the exchange energy $\Delta_{ex} = E_{AFM} - E_{FM}$. We consider only the hopping ($t$) between the $p$-orbital of L and the $d$-orbital of TM, since two transition-metal atoms are far apart. The numerically calculated exchange energy $\Delta_{ex}$ was plotted as a function of the hopping parameter $t$ in the left panel of FIG. 3, which indicates that the ground state is AFM. The same conclusion can be reached using the perturbation theory (for details, see Section VI of the SM), in agreement with the Goodenough-Kanamori rule.

Now we consider a four-site cluster (TM-L(M)-TM) model derived from the three-site cluster model by attaching an atom (M) with one empty $p$-orbital to L (shown in the right panel of FIG. 3). In our case, Bi takes the role of the atom M. In addition to the hopping $t$ from the $p$-orbital of L to the $d$-orbital of TM, there exists another hopping $t'$ from the $p$-orbital of M to the $p$-orbital L. Note that the hybridization between TM and M is omitted, since they are far apart. At this stage, we discuss the occupied energy in two different magnetic configurations, FM and AFM. **(a) FM alignment**. Due to the charge balance, the three spin-up electrons must occupy three spin-up orbitals of the two TM atoms and one L atom. In the spin-up manifold, the $p$-$d$ hopping pushes up the $p$-orbital of L by $2t^2/\Delta_{pd}$. Here, $\Delta_{pd}= \varepsilon_p - \varepsilon_d$ is the on-site energy difference between $d$-orbital of TM and the $p$-orbital of L, which is denoted in FIG. 4(a). Note that in our case, $\Delta_{pd}> 0$. The introduction of M atom pushes down the $p$-orbital of L by $\delta^{\uparrow}_{p'p} = \frac{t'^2}{\Delta_{p'p}-2t^2/\Delta_{pd}}$, where $\Delta_{p'p}= \varepsilon_{p'} - \varepsilon_p$ is the on-site energy difference between the $p$-orbital of M and that of L. To keep the orbital of M atom mostly empty, we requires $\frac{\Delta_{p'p}\Delta_{pd}}{t^2} > 2$. For the spin-down manifold, we performed a similar analysis by noting that the $p$-orbital of L is pushed down by $2t^2/(U - \Delta_{pd})$. Here, $U$ is the on-site repulsion of the TM atom, which is denoted in FIG. 4. By introducing the surrounding atom M, the $p$-$p$ hopping pushes down the $p$-orbital of L by $\delta^{\downarrow}_{p'p} = \frac{t'^2}{\Delta_{p'p}+2t^2/(U-\Delta_{pd})}$, as shown in FIG. 4(b). Note that the spin-down electron occupies the $p$-orbital of L. The total energy gain for the FM alignment due to the introduction of M is given by: $\Delta E_{FM} = \delta^{\uparrow}_{p'p} + \delta^{\downarrow}_{p'p}$. **(b) AFM alignment**. In this case, two spin-up electrons occupy the $d$-orbital of the left TM and the $p$-orbital of L, while two spin-down electrons occupy $d$-orbital of the right TM and the $p$-orbital of L. We first consider the spin-up manifold without the atom M. Then, the high energy empty $d$-orbital of TM hybridizes with the $p$-orbital of L to push the latter by $t^2/(U - \Delta_{pd})$. The hybridization between the two occupied orbitals also pushes up the $p$-orbital of L by $t^2/\Delta_{pd}$. Hence the $p$-orbital of L shifts by $t^2/\Delta_{pd} - t^2/(U - \Delta_{pd})$. Now we include the influence of the atom M. The $p$-$p$ hopping between the $p$-orbitals of M and L pushes down the $p$-orbital of L by $\delta^{\uparrow}_{p'p} = \frac{t'^2}{\Delta_{p'p} - t^2/\Delta_{pd} + t^2/(U - \Delta_{pd})}$, as shown in FIG. 4(b). For the spin-down manifold, the energy gain is same as the spin-up manifold. Therefore, the total energy gain due to the introduction of M is given by: $\Delta E_{AFM} = 2\delta^{\uparrow}_{p'p}$. Comparing the energy gains in the FM and AFM states caused by introducing the surrounding atom M, we obtain $\Delta E_{gain} = \Delta E_{FM} - \Delta E_{AFM}$:

$$\Delta E_{gain} = \frac{2\alpha^2 t'^2 \Delta_{pd}}{t^2(\beta - 2)[(\alpha - 1)\beta + 2][(\alpha - 1)(\beta - 1) + 1]}$$

where $\alpha = \frac{U}{\Delta_{pd}} > 1$ in order to keep the L orbital occupied, and $\beta = \frac{\Delta_{p'p}\Delta_{pd}}{t^2} > 2$ as mentioned before. Therefore, $\Delta E_{gain}$ is always larger than 0, so the energy gain of FM state is greater than that of AFM alignment. This mechanism can also be intuitively seen from FIG.4(a) and (c). In FM state, the $p$-$p$ hybridization is stronger than that in AFM, which results in a greater lowering of the $p$-orbital of L. The difference in the different energy gains is proportional to the square of the $p$-$p$ hopping $t'$. Therefore, the ground state of the system depends on the competition between the different energy gains, $\Delta E_{gain}$, and the exchange energy, $\Delta_{ex}$, of the three-site cluster. The stronger the $p$-$p$ hopping, the more likely it is to achieve the FM ground state. In FIG.3(b), we plot the numerically computed exchange energy as a function of the $p$-$p$ hopping $t'$ to find that an AFM-to-FM transition can indeed occur when $|t'|$ is increasing.

As discussed above and in Section VI of the SM, each $Bi^{3+}$ cation attached to the inner $Te^{2-}$ anion forming the Mn-Te-Mn superexchange path is crucial for the FM coupling between the $Mn^{2+}$ magnetic ions. Additionally, as mentioned earlier, SOC enhances ferromagnetism. Comparing the PDOS without and with SOC (see Section IV of the SM), we find that the Bi $6p$-orbital is lowered in energy under SOC. The latter will enhance the $p$-$p$ hybridization between the $p$-orbitals of M and L, resulting in a stronger FM coupling between the two TM. This conclusion is consistent with our four-sites model.

To summarize, the origin of the FM coupling in $MnBi_2Te_4$ was explored in three steps. Firstly, we confirmed the FM state is the ground state of a $MnBi_2Te_4$ monolayer through DFT calculations. Then, we found that the $Bi^{3+}$ cation attached to the inner $Te^{2-}$ anion forming the Mn-Te-Mn superexchange path plays a crucial role for the FM coupling between the adjacent $Mn^{2+}$ cations through performing test DFT calculations. Finally, using a model TB Hamiltonian, it was shown why the hybridization of the Bi-$6p$ orbital with the $5p$ orbital of the inner Te makes the FM coupling more favorable energetically than the AFM coupling. The mechanism of the FM coupling in $MnBi_2Te_4$ revealed in this work is fundamentally different from that in $CrI_3$ and $CrGeTe_3$ where the coupling between the occupied $t_{2g}$ orbital and empty $e_g$ orbital is the key.

**Acknowledgments.** This work is supported by NSFC (No. 11825403), Program for Professor of Special Appointment (Eastern Scholar), Qing Nian Ba Jian Program. J.F. acknowledges the support from Anhui Provincial Natural Science Foundation (1908085MA10). H.-J.K. thanks the support of the Basic Sciences Research Program through the National Research Foundation of Korea (NRF) funded by the Ministry of Education (NRF-2017R1D1A1B03029624). H. X. Thank Dr. Jing Wang for useful discussion.

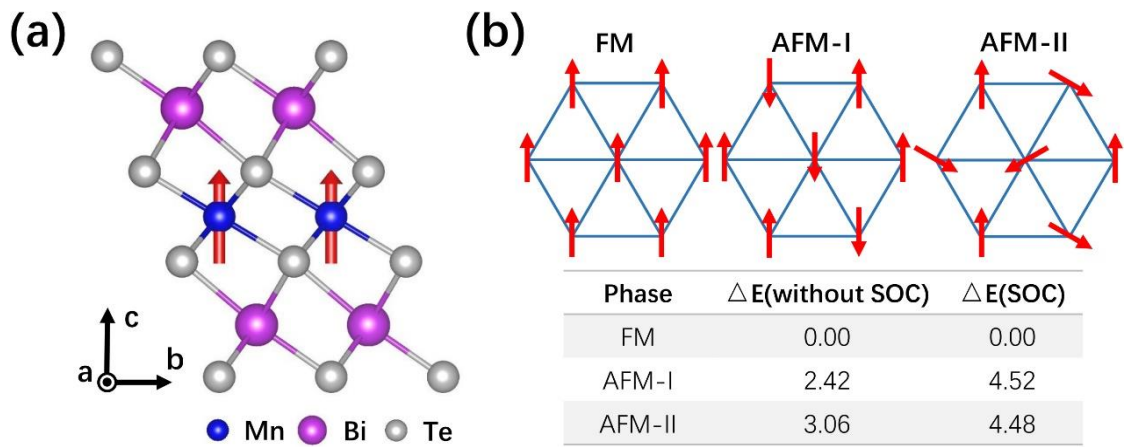

**FIG. 1.** (a) Side view of a $MnBi_2Te_4$ monolayer consisting three sheets of cations and four sheets of anions. (b) The total energy of $MnBi_2Te_4$ in different magnetic states by setting the total energy of FM as reference in units of meV/f.u.. Blue triangular lattice

represents the Mn layer and the red arrows represent the spin moments. There are three kinds of spin orders: FM (the upper left plane), AFM-I (AFM coupling between FM chains as shown in the upper center), AFM-II (120° angle between any adjacent spin in the upper right plane). The lower panel of (b) lists the exchange energy without and with SOC for FM, AFM-I and AFM-II.

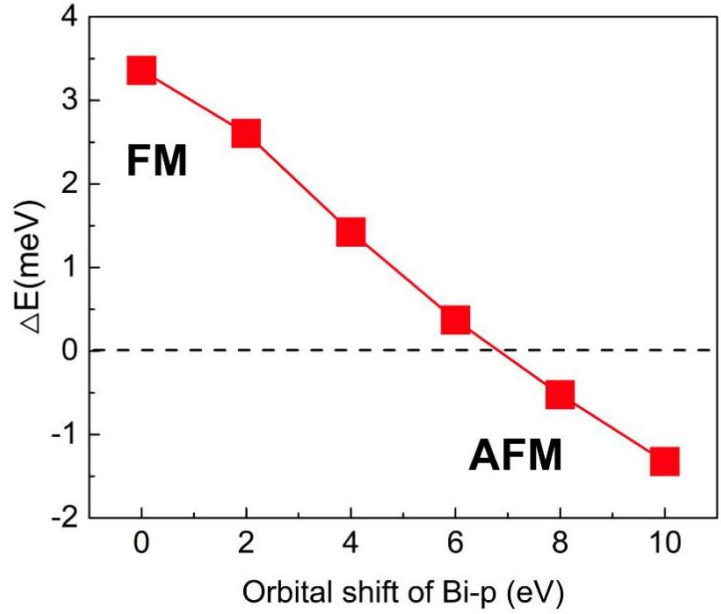

**FIG. 2.** The exchange energy $\Delta_{ex} = E_{AFM} - E_{FM}$ as a function of the energy shift of the Bi-$6p$ orbitals.

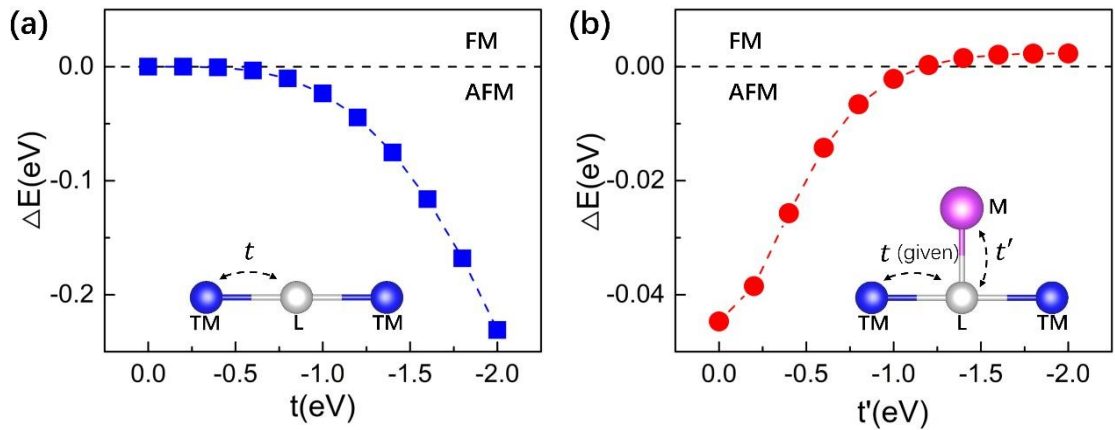

**FIG. 3.** The exchange energies obtained for the three-site and four-site cluster models. The left panel shows the exchange energy of the three-site cluster model (inset) as a function of the hopping parameter $t$. The right panel shows the exchange energy of the four-site cluster model as a function of the hopping parameter $t'$ with the hopping $t$ fixed to a reasonable value (-1.2 eV). In both cases, the onsite energy is set according to the PDOS of MnBi$_2$Te$_4$.

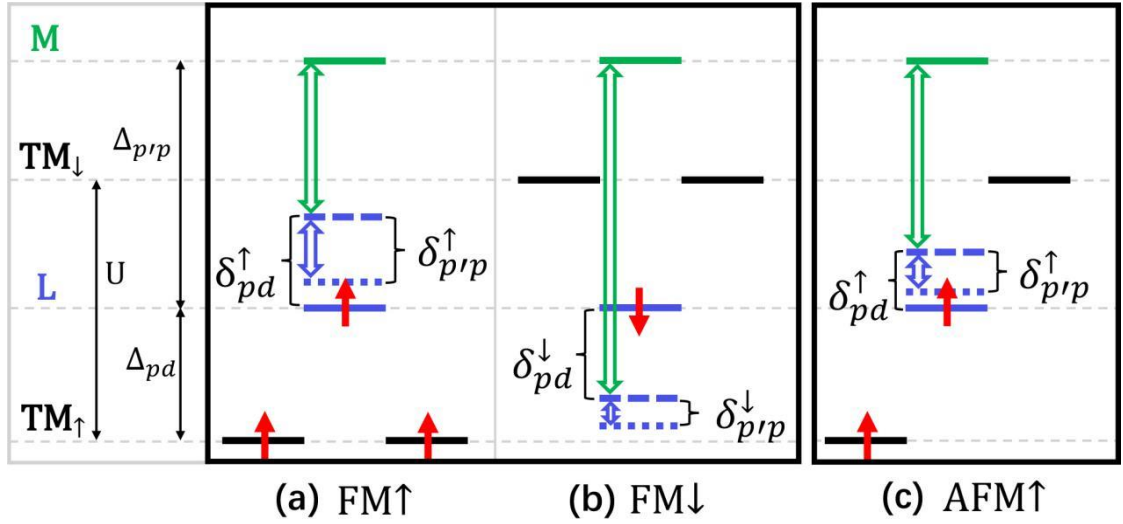

**FIG. 4.** Schematic diagram for the FM mechanism in MnBi$_2$Te$_4$. (a) and (b) show the up-spin and down-spin energy levels in the case of FM coupling between the two TM ions, respectively. (c) illustrates the up-spin energy levels in the case of AFM coupling between the two TM ions (the down-spin levels are same to the up-spin case). The blue dashed lines refer to the ligand levels after the *p-d* hybridization between the TM and L atoms, and the blue solid lines represent the ligand levels after the *p-p* hybridization between the L and M atoms. The green hollow double-headed arrows mark the energy difference between the L and M atoms after the *p-d* hybridization. The blue hollow double-headed arrows represent the energy gained from the *p-p* hybridization, and the red arrows the electron with definite spin state.

**References**

[1] M. Z. Hasan and C. L. Kane, Rev. Mod. Phys. **82**, 3045 (2010).

[2] X.-L. Qi and S.-C. Zhang, Rev. Mod. Phys. **83**, 1057 (2011).

[3] B. A. Bernevig, T. L. Hughes, and S.-C. Zhang, Science **314**, 1757 (2006).

[4] R. Yu, W. Zhang, H.-J. Zhang, S.-C. Zhang, X. Dai, and Z. Fang, Science **329**, 61 (2010).

[5] H. Pan, Z. Li, C.-C. Liu, G. Zhu, Z. Qiao, and Y. Yao, Phys. Rev. Lett. **112**, 106802 (2014).

[6] Z. Wang, Z. Liu, and F. Liu, Phys. Rev. Lett. **110**, 196801 (2013).

[7] H. Zhang *et al.*, Phys. Rev. Lett. **108**, 056802 (2012).

[8] Z. Qiao, W. Ren, H. Chen, L. Bellaiche, Z. Zhang, A. H. MacDonald, and Q. Niu, Phys. Rev. Lett. **112**, 116404 (2014).

[9] J. Wang, B. Lian, and S.-C. Zhang, Phys. Scr. **2015**, 014003 (2015).

[10] C.-Z. Chang *et al.*, Science **340**, 167 (2013).

[11] Q. L. He *et al.*, Nat. Mater. **16**, 94 (2017).

[12] M. Mogi, M. Kawamura, R. Yoshimi, A. Tsukazaki, Y. Kozuka, N. Shirakawa, K. Takahashi, M. Kawasaki, and Y. Tokura, Nat. Mater. **16**, 516 (2017).

[13] L. A. Wray *et al.*, Nat. Phys. **7**, 32 (2011).

[14] D. Xiao *et al.*, Phys. Rev. Lett. **120**, 056801 (2018).

[15] E. Rienks *et al.*, Nature **576**, 423 (2019).

[16] D. Zhang, M. Shi, T. Zhu, D. Xing, H. Zhang, and J. Wang, Phys. Rev. Lett. **122**, 206401 (2019).

[17] J. Li, Y. Li, S. Du, Z. Wang, B.-L. Gu, S.-C. Zhang, K. He, W. Duan, and Y. Xu, Sci. Adv. **5**, eaaw5685 (2019).

[18] Y. Deng, Y. Yu, M. Z. Shi, J. Wang, X. H. Chen, and Y. Zhang, arXiv preprint arXiv:1904.11468 (2019).

[19] Y. Gong *et al.*, Chin. Phys. Lett. **36**, 076801 (2019).

[20] M. Otrokov *et al.*, Nature **576**, 416 (2019).

[21] R. S. Mong, A. M. Essin, and J. E. Moore, Phys. Rev. B **81**, 245209 (2010).

[22] X.-L. Qi, T. L. Hughes, and S.-C. Zhang, Phys. Rev. B **78**, 195424 (2008).

[23] A. M. Essin, J. E. Moore, and D. Vanderbilt, Phys. Rev. Lett. **102**, 146805 (2009).

[24] F. Wilczek, Phys. Rev. Lett. **58**, 1799 (1987).

[25] C. Gong *et al.*, Nature **546**, 265 (2017).

[26] B. Huang *et al.*, Nature **546**, 270 (2017).

[27] J. B. Goodenough, *Magnetism and chemical bond* (Interscience Publ., 1963), Vol. 1.

[28] J. Kanamori, J. Phys. Chem. Solids **10**, 87 (1959).

[29] P. Anderson, Phys. Rev. **79**, 350 (1950).

[30] J. Kanamori, Prog. Theor. Phys. **17**, 177 (1957).

[31] J. B. Goodenough and A. L. Loeb, Phys. Rev. **98**, 391 (1955).

[32] J.-Q. Yan *et al.*, Phys. Rev. Mater. **3**, 064202 (2019).

[33] H. Xiang, C. Lee, H.-J. Koo, X. Gong, and M.-H. Whangbo, Dalton Trans. **42**, 823 (2013).

[34] H. Xiang, E. Kan, S.-H. Wei, M.-H. Whangbo, and X. Gong, Phys. Rev. B **84**, 224429 (2011).

[35] M. Freiser, Phys. Rev. **123**, 2003 (1961).

[36] Y. Miyatake, M. Yamamoto, J. Kim, M. Toyonaga, and O. Nagai, J. Phys. C: Solid State Phys. **19**, 2539 (1986).

[37] X. Wan, J. Zhou, and J. Dong, EPL (Europhysics Letters) **92**, 57007 (2010).

[38] Y. Du, H.-C. Ding, L. Sheng, S. Y. Savrasov, X. Wan, and C.-G. Duan, J. Phys.: Condens. Matter **26**, 025503 (2013).

[39] J. Feng and H. Xiang, Phys. Rev. B **93**, 174416 (2016).

[40] H. Katsura, N. Nagaosa, and A. V. Balatsky, Phys. Rev. Lett. **95**, 057205 (2005).